# Hardy-type Nonlocality Proof for Two Maximally Entangled Particles

## Demetrios Kalamidas

*Institute for Ultrafast Spectroscopy and Lasers, City College of the City University of New York, 138[th] Street & Convent Avenue, New York, NY, 10031, USA*

**Abstract**

We present a variation on a gedanken experiment of Hardy [Phys. Rev. Lett. 68 (1992) 2981] that allows a Hardy-type nonlocality proof for two maximally entangled particles in a four-dimensional Hilbert space.

## 1.    Introduction

By way of an ingenious gedanken experiment, Hardy [1] constructed a simple proof of Bell's theorem [2] without inequalities for two entangled particles in a four-dimensional total Hilbert space. However, the two-particle entangled state obtained in the gedanken experiment of [1] was not a maximally entangled state. Subsequently, (in [3]) Hardy went on to show that his nonlocality proof can only go through for nonmaximally entangled states of two particles. In this Letter we present a variation on his gedanken experiment that does allow a Hardy-type nonlocality argument to be developed for two particles in a maximally entangled Bell-state.

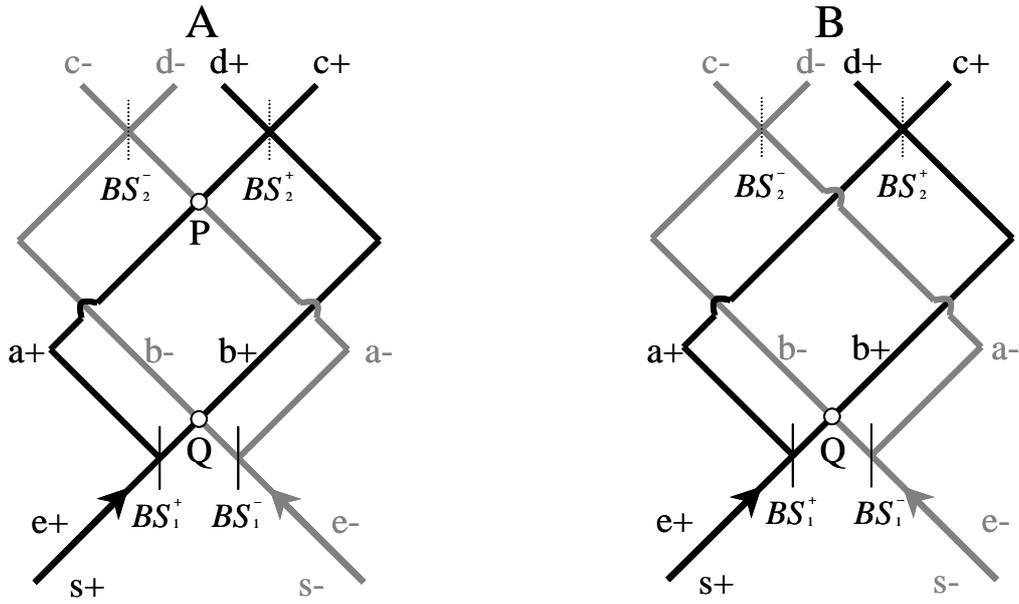

Fig.1. Based on Hardy's gedanken experiment, the above configurations are used to demonstrate a contradiction between local realism and quantum mechanics. In 'A' there are two points of intersection (at P and Q) while in 'B' there is only one (at Q).

## 2. Gedanken Experiment

Consider scheme 'A' of Fig.1. A positron $e^+$ and an electron $e^-$ are created at distant locations and each particle enters its own interferometer via paths $s^+$ and $s^-$, respectively. The two particles simultaneously reach their corresponding beam splitters, $BS_1^{\pm}$, which transform their inputs as follows:

$$|s^{\pm}\rangle \to \frac{1}{\sqrt{2}}(|b^{\pm}\rangle + i|a^{\pm}\rangle),$$

where the kets signify the presence of a particle propagating along a certain path (i.e. $|b^+\rangle$ represents a positron in path b$^+$). Accordingly, beam splitters $BS_2^\pm$ transform their inputs as follows:

$$|a^\pm\rangle \to \frac{1}{\sqrt{2}}(|c^\pm\rangle + i|d^\pm\rangle), \qquad |b^\pm\rangle \to \frac{1}{\sqrt{2}}(|d^\pm\rangle + i|c^\pm\rangle).$$

Beam splitters $BS_1^\pm$ are fixed in their place for *all* experiments while beam splitters $BS_2^\pm$ can be individually removed at the experimenter's pleasure. Paths b$^+$ and b$^-$ intersect at point Q while paths a$^+$ and a$^-$ intersect at point P. Hence, if the positron-electron pair meets at Q or P by having occupied one or the other of these path combinations, it annihilates into a photon. The state of the photon is denoted by $|\gamma\rangle^{P/Q}$, depending on where it was created.

Suppose beam splitters $BS_2^\pm$ are removed. We will use the notation $E$(P,Q; +out,-out), indicating the points of intersection and whether each of these beam splitters is 'in' or 'out'. For this experiment, the state of the positron-electron pair evolves as

$$|s^+\rangle|s^-\rangle \to \frac{1}{2}(|b^+\rangle + i|a^+\rangle)(|b^-\rangle + i|a^-\rangle) \to$$
$$\frac{1}{2}|\gamma\rangle^Q - \frac{1}{2}|\gamma\rangle^P + \frac{i}{2}(|d^+c^-\rangle + |c^+d^-\rangle). \qquad (1)$$

Let us now examine scheme 'B' of Fig.1. In this configuration there is only one point of intersection (at Q). We will consider two experiments with this configuration: For experiment $E(Q; +\text{in},-\text{out})$ the final state of the positron-electron pair is given by

$$\frac{1}{2}|\gamma\rangle^Q + \frac{1}{2\sqrt{2}}(-2|c^+c^-\rangle + i|c^+d^-\rangle - |d^+d^-\rangle) \quad (2)$$

while for experiment $E(Q; +\text{out},-\text{in})$ the final state is given by

$$\frac{1}{2}|\gamma\rangle^Q + \frac{1}{2\sqrt{2}}(-2|c^+c^-\rangle + i|d^+c^-\rangle - |d^+d^-\rangle). \quad (3)$$

Finally, returning to configuration 'A', we will consider experiment $E(P,Q; +\text{in},-\text{in})$, for which the final state is given by

$$\frac{1}{2}|\gamma\rangle^Q - \frac{1}{2}|\gamma\rangle^P - \frac{1}{2}(|c^+c^-\rangle + |d^+d^-\rangle). \quad (4)$$

We will now demonstrate how the above set of experiments, within the framework of a Hardy-type proof, leads to a contradiction between local realism and certain predictions of quantum mechanics. Let us first prescribe what is meant by local realism in the context of the above experiments:

By invoking realism we assume that the positron-electron pair is completely described by some set of hidden variables, $\{\lambda\}$. The variables comprising $\{\lambda\}$ may take on different values for each pair of an ensemble of what quantum mechanics regards as identically prepared states. When formulated in terms of hidden variables, realism requires that the outcomes of any measurements performed on each particle of the positron-electron pair have been predetermined by the values within $\{\lambda\}$. This, in turn, demands that the outcomes inferred from one experimental context remain fixed when counterfactual reasoning is used to determine outcomes in another experimental context. By invoking locality we require that the outcome of a measurement on one particle does not depend on the kinds of measurements performed on other space-like separated particles.

Suppose there is a detector in each of the four output paths $c^+$, $d^+$, $c^-$, $d^-$. If, for instance, a particle is registered by the detector in path $d^+$, with beam splitter $BS_2^+$ in place, we will write $D_\lambda^+(in) = 1$; if the particle is not detected we will write $D_\lambda^+(in) = 0$. We follow this notation for the rest of the detectors. Note that the dependence of outcomes upon the hidden variables $\lambda$ is now explicitly shown.

From experiment $E(P,Q; +\text{out},-\text{out})$ we have

$$D_\lambda^+(out)D_\lambda^-(out) = 0 \qquad (5)$$

for *all* trials because there is no $|d^+d^-\rangle$ term in (1). Had we performed experiment $E(Q;+\text{in},-\text{out})$, we would have the implication

$$\text{If} \quad D_\lambda^+(in) = 1 \quad \text{then} \quad D_\lambda^-(out) = 1 \qquad (6)$$

because $|d^+\rangle$ only appears in the term $-|d^+d^-\rangle$ of (2). Similarly, had we performed experiment $E(Q;+out,-in)$, we would have the implication

$$\text{If} \quad D_\lambda^-(in) = 1 \quad \text{then} \quad D_\lambda^+(out) = 1 \qquad (7)$$

because $|d^-\rangle$ only appears in the term $-|d^+d^-\rangle$ of (3). Lastly, had we performed experiment $E(P,Q;+in,-in)$, we would have

$$D_\lambda^+(in)D_\lambda^-(in) = 1 \qquad (8)$$

for 25% of trials because of the term $-\frac{1}{2}|d^+d^-\rangle$ contained in (4). However, from (6) and (7) we have that $D_\lambda^+(in)D_\lambda^-(in) = 1$ necessarily implies that $D_\lambda^+(out)D_\lambda^-(out) = 1$ which, in turn, contradicts (5) since there it was determined that the value of this product of outcomes could never be equal to 1. This contradiction shows that a local realistic description of quantum mechanical predictions, in terms of local hidden variables, is untenable for the above set of experiments.

## 3. Conclusions

Upon inspection of (1) we realize that the state of the positron-electron pair obtained when the two particles survive annihilation is, in fact, a *maximally entangled state*: we can re-write $\frac{i}{2}(|d^+c^-\rangle + |c^+d^-\rangle)$ as $\frac{i}{\sqrt{2}}|\Psi\rangle$, where $|\Psi\rangle = \frac{1}{\sqrt{2}}(|d^+c^-\rangle + |c^+d^-\rangle)$ is a Bell-state. Similarly, the final state in (4) is $\frac{-1}{\sqrt{2}}|\Phi\rangle$, where $|\Phi\rangle = \frac{1}{\sqrt{2}}(|c^+c^-\rangle + |d^+d^-\rangle)$ is another Bell-state. Therefore, the contradiction between (5) and (8) demonstrates that a Hardy-type proof of nonlocality *is* possible for Bell-states, contrary to what is currently believed [3,4]. Furthermore, we have seen that the result $D_\lambda^+(in)D_\lambda^-(in) = 1$ is obtained in 25% of trials, exceeding the maximum of $\approx 9\%$ derived by Hardy [3] for nonmaximally entangled states. However, like Hardy's original proposal [1], the idealized experiment outlined here is only possible 'in principle' and it remains to be seen whether an experimentally feasible (i.e. quantum optical) analogue can be found. An interesting attempt was made by Wu *et al.* [5] to demonstrate a Hardy-type contradiction between local realism and quantum mechanics for two photons created in a maximally entangled state. They proposed an experimentally feasible interferometric set-up that was a variation on what Horne *et al.* presented in [6]. However, their claims have been rightfully criticized, we believe, by Cereceda [7] who notes that the required minimum dimensionality of the total Hilbert space is six instead of four. The arguments of Wu *et al.* cannot go through in four dimensions, whereas Hardy's demonstration requires only four dimensions. To our knowledge, this Letter outlines, for the first time, a Hardy-type nonlocality proof for two particles in a maximally entangled state in a four-dimensional total Hilbert space.

## 4.  Acknowledgements

I am very grateful to D. M. Greenberger for discussions and guidance, and to R. R Alfano for the use of his facilities at IUSL. Partial financial support for this research was provided by NASA.